\documentclass[12pt]{article}
\usepackage{cite}
\usepackage{amssymb}
\usepackage{amsmath}
\usepackage{epsf,graphicx}

\newcommand\ZZ{\hbox{\zfont Z\kern-.4emZ}}
\font\zfont = cmss10 

\newcommand{\lam}{{\lambda}}

\begin{document}

\begin{titlepage}
\begin{flushright}
{\tt hep-th/0301130} \\
\end{flushright}

\vskip.5cm
\begin{center}
{\huge{\bf Holographic Weyl Entropy Bounds}}
\vskip.2cm
\end{center}
\vskip0.2cm

\begin{center}
{\sc Andrew Chamblin}$^{a}$ and
{\sc Joshua Erlich}$^{b}$ \\

\end{center}
\vskip 10pt

\begin{center}
\vspace*{0.1cm}
$^{a}$ {\it Department of Physics, Queen Mary, University of London, Mile End Road, 
London E1 4NS }\\
$^{b}$ {\it Department of Physics, Box 1560, University of Washington, Seattle, WA
98195, USA } \\

\vspace*{0.1cm}
{\tt chamblin@mit.edu, erlich@phys.washington.edu}
\end{center}

\vglue 0.3truecm

\begin{abstract}
\vskip 3pt \noindent 
We consider the entropy bounds recently conjectured by Fischler, Susskind and
Bousso, and proven in certain cases by Flanagan, Marolf and Wald 
(FMW).  One of the FMW derivations supposes a covariant form of the Bekenstein
entropy bound, the consequences of which we explore.  The derivation also
suggests that the entropy
contained in a vacuum spacetime, e.g.  Schwarzschild, is related to
the shear on congruences of null rays.  We find evidence for this intuition, 
but in a surprising
way.  We compare the covariant entropy bound
to certain earlier discussions of black hole entropy, and comment on the
separate roles of quantum mechanics and gravity in the entropy bound.

\end{abstract}

\end{titlepage}

\newpage


\section{Introduction}
\label{sec:intro}
\setcounter{equation}{0}
\setcounter{footnote}{0}

Various authors have put forward the idea
that a `Holographic Principle' should be incorporated into any
attempt to construct a quantum theory of gravity.  This principle, which was
first developed in papers by 't Hooft \cite{thooft} and Susskind \cite{lenny},
is on the surface a radical statement about 
how many degrees of freedom there are in
Nature.  In essence, the principle asserts that a physical system can be
completely described by information which is stored at the boundary of the
system, without exceeding one bit of information per unit Planck area.
Much study has been devoted to the related
topics of black holes, entropy bounds, and holography, and
we will not be able to do justice to the bulk of prior work in the field.  For reviews
see, for example, \cite{waldrev,lennyrev,smolinrev,boussorev} and references therein.

For some time, there was no precise covariant statement of the Holographic Principle;
however, this situation was rectified in a series of elegant papers by
Fischler and Susskind \cite{fs} and
Bousso \cite{raph,raph1,raph2}.  In particular, by
choosing appropriate lightlike surfaces (called `lightsheets') 
where the entropy of a given
system can reside, Bousso was able to develop a 
mathematically precise covariant entropy conjecture.  
Bousso defined a lightsheet $\Gamma$ associated with a spacelike two-surface
$B$ as a null congruence orthogonal to $B$.  
The lightsheet is
terminated at caustics, spacetime boundaries and singularities.  
A Bousso lightsheet has the further property that the expansion $\theta$ 
(see \cite{wald}, for example) is everywhere nonpositive.  
The covariant entropy
bound is then the statement that the entropy contained in a Bousso lightsheet 
is bounded by the area of
its initial boundary $B$, or simply (suppressing $\hbar$ and Newton's constant
$G$), \begin{equation}
S(\Gamma)\leq A(B)/4. \label{eq:Bousso}\end{equation}

Soon after the work of Fischler, Susskind and Bousso (FSB), a proof
of various classical versions of Bousso's bound was provided by
Flanagan, Marolf and Wald (FMW) \cite{fmw}.  In order to make a mathematically
precise statement, which could consequently be proven, they took the
step of introducing an entropy flux vector,
denoted $s^a$.  The total entropy through a given lightsheet $\Gamma$ is then
defined to be the integral of $s^a$ over the surface of the lightsheet.
They showed that FSB-type bounds could be proven,
provided the entropy flux vector satisfied one of the following two sets of
criteria:

Either,
\begin{equation}
s_\Gamma\cdot k\leq (1-\lambda) (\pi\, k_a T^{ab}k_b + \sigma_{ab}\sigma^{ab}/8),
\label{eq:fmw1}
\end{equation}
or
\begin{equation}\left\{\begin{array}{l}
(s_a k^a)^2 \le T_{ab} k^a k^b / (16 \pi) + {\sigma_{ab}}{\sigma^{ab}} / 
(128 {\pi}^{2}), \\  \\
|k^a k^b \nabla_a s_b| \le \pi T_{ab} k^a k^b /4 + {\sigma_{ab}}{\sigma^{ab}} /
 32, \end{array} \right.\label{eq:fmw2}
\end{equation}
\noindent where $k^a$ denotes the tangent vector to a given null geodesic 
generating the light sheet in question, $T_{ab}$ denotes the stress-energy
tensor, and $\sigma_{ab}$ denotes the shear tensor of the null congruence 
\cite{he}.  The affine parameter $\lambda$ is normalized to
range from 0 to 1 over the lightsheet.
The first condition (\ref{eq:fmw1}) is defined for each lightsheet $\Gamma$,
while the second set of conditions (\ref{eq:fmw2}) are 
defined pointwise.  The condition (\ref{eq:fmw1}) is reminiscent of the
Bekenstein bound, \begin{equation}
S\leq 2\pi ER, \label{eq:Bekenstein} \end{equation}
where $E$ is the energy contained in a system of size $R$.  
(Reference \cite{Bousso-Bek} also discusses the Bekenstein bound and its relation
to the Bousso bound.  In particular, a version of the Bekenstein bound is derived from
the Bousso bound, reversing the logic discussed here.)
We will refer
to the relation (\ref{eq:fmw1}) as the covariant Bekenstein bound. 
Indeed, we will
note in the next section
that for spherically symmetric systems the condition (\ref{eq:fmw1}) is
quite similar to (\ref{eq:Bekenstein}).  A stronger form of the Bousso
bound follows from (\ref{eq:fmw1}), but not (\ref{eq:fmw2}).  If the lightsheet
$\Gamma$ is terminated on some spacelike 2-surface $B'$, then given (\ref{eq:fmw1})
the entropy
in the lightsheet was shown in \cite{fmw} to satisfy, \begin{equation}
S\leq \frac{A(B)-A(B')}{4}. \label{eq:Bousso2}\end{equation}
In fact, we will find that this bound can never be saturated unless $A(B')=0$.

Flanagan, Marolf and Wald chose to consider somewhat stronger conditions on
the entropy flux vector than (\ref{eq:fmw1}) or (\ref{eq:fmw2})
by suppressing the terms proportional to
$\sigma_{ab}^2$, although the neglected terms can be included 
without violating the Bousso bound.
It is suggested in \cite{fmw} that the neglected terms
might be related to gravitational
entropy, an idea which we will explore further in this paper.  The possible
relation betwen the shearing of null congruences and gravitational entropy
is reminiscent of Penrose's suggestion that the shear should
be interpreted as a measure of gravitational {\em energy} \cite{rogerclassic}. 
The Weyl tensor acts as a source for the shear tensor, and hence the shearing of
null rays can naturally be interpreted as due to a 
gravitational contribution to the energy.
Amusingly, some of the ideas in Penrose's paper are tantalizingly 
similar to those of Fischler, Susskind and Bousso if one replaces the 
concept of entropy with energy.
Indeed, quoting from
\cite{rogerclassic}:
\vspace*{0.5cm}

\noindent {\it ``..it is suggested that the resultant focusing power of
spacetime curvature along a null ray is a good measure of the total energy
flux (matter plus gravitation) across the ray ..''}

\vspace*{0.5cm}
In other words, Penrose was interested in the idea of measuring the total
energy (as opposed to entropy) which can flow through a given lightsheet.
Generically there are two  different types
of focusing which may occur along a Bousso lightsheet: anastigmatic focusing 
due to the stress-energy tensor $T_{ab}$,
and purely astigmatic focusing due to the shear tensor $\sigma_{ab}$.  According
to Penrose, the latter
is to be interpreted as due to 
gravitational energy, and indeed a relation similar to (\ref{eq:Bekenstein}) 
would then imply a relation between the shear and gravitational entropy.
The intuition that the Weyl tensor should somehow count gravitational degrees
of freedom is also reflected in \cite{hp}.


The physical interpretation of the entropy flux vector is somewhat obscure, and
in general it is certainly not clear that a quasi-local entropy current should
exist.  However, if an entropy flux vector can be suitably defined, both sets of 
conditions (\ref{eq:fmw1}) and (\ref{eq:fmw2}) are reasonable under a large 
class of situations \cite{fmw}.  It should also be noted that these 
conditions could be violated in situations in which the number of species
of matter is large enough.  Hence, by assuming either of the above sets of
conditions the species problem is swept under the rug.  
Be that as it may, in the hope that it will indeed follow from a fundamental 
theory including gravity, at least in a large class of situations,
we would like to take the Bekenstein-like relation
(\ref{eq:fmw1}) seriously and further explore its consequences.

In particular, we will investigate how 
the shear tensor of a given lightsheet might be a measure of
the number of gravitational degrees of freedom.  On the surface,
this seems counterintuitive because many null geodesic congruences will have
vanishing shear, even though the spacetime may have large curvature.
Indeed, spherically symmetric lightsheets in
Schwarzschild spacetime are shear-free.  
As a result, the potential relationship between shear and entropy is
{\em a priori} dubious, and leads us to consider less symmetric Bousso
lightsheets.  We introduce the concept of a maximal entropy lightsheet,
and are led to a type of ultraviolet-infrared duality between matter and
gravitational entropy.

\section{Consequences of the covariant Bekenstein bound}
In this section we will consider the properties of the covariant Bekenstein
bound (\ref{eq:fmw1}) and its relation to the Bousso bound.  Flanagan,
Marolf and Wald \cite{fmw} 
derived the Bousso bound (\ref{eq:Bousso}) from the covariant Bekenstein bound 
(\ref{eq:fmw1}).
It is interesting that saturation of the inequality (\ref{eq:fmw1})
does not imply saturation of the Bousso bound (\ref{eq:Bousso}).
We begin by studying the conditions for saturation of the Bousso bound.  Black holes
are expected to saturate ``useful'' entropy bounds, and as
black holes provide the motivation for most of these ideas,
we would like to learn what we can about them
from reasonable assumptions like the covariant Bekenstein bound.  To this end, we
first review the derivation of the Bousso bound from (\ref{eq:fmw1}),
following \cite{fmw}.

The area factor ${\cal A}(\lam)$ is given by, \begin{equation}
{\cal A}(\lam)=\exp \int_0^\lam d\overline{\lam}\,\theta(\overline{\lam}),
\end{equation}
where $\theta$ 
is the expansion along a null ray in the congruence
(see, for example, \cite{wald}).  ${\cal A}(\lam)$ measures the ratio of the area
of the spacelike slice of the lightsheet at affine parameter $\lambda$ to the
area of the boundary 2-surface, $A(B)$.  It is helpful
to define, as in \cite{fmw}, \begin{equation}
G(\lam)=\sqrt{{\cal A}(\lam)}.\end{equation}
The entropy $S(\Gamma)$ is given by \cite{fmw}: \begin{equation}
S(\Gamma)=A(B)\int_0^1d\lam\,s_ak^a\,{\cal A}(\lam). \end{equation} 
From this and (\ref{eq:fmw1}), the generalized Bousso bound 
(\ref{eq:Bousso2}) is equivalent to the statement that
along each null ray $k$ in the congruence, 
\begin{equation}
I_\gamma \equiv \frac{1}{8}\int_0^1 d\lam\,(1-\lam)\,(\sigma^2 + 8\pi\, kTk)
\,{\cal A}(\lam)
< \frac{1}{4}(1-{\cal A}(1)).
\label{eq:fmwproof}\end{equation}
Using
the Raychaudhuri equation,
\begin{equation}
-\frac{d\theta}{d\lam}=\frac{1}{2}\theta^2+ 8\pi \, k_a T^{ab}k_b + 
\sigma_{ab}^2, \label{eq:Ray} \end{equation}
and integrating by parts,
we can rewrite $I_\gamma$ as,
\begin{eqnarray}
I_\gamma&=&-\frac{1}{4}d\lam\,(1-\lam)\,G''(\lam)\,G(\lam) \\
&=& -\frac{1}{4}\int_0^1 d\lam\,(1-\lam)\,G''(\lam)+\frac{1}{4}\int_0^1 d\lam\,
(1-\lam)\,G''(\lam)(1-G(\lam)) \\
&=&\frac{1}{4}\left[ G(0)-G(1)+G'(0) \right]+\frac{1}{4}\int_0^1 d\lam\,(1-\lam)\,
G''(\lam)\,(1-G(\lam)). \end{eqnarray}
But $G(0)=1$ and $G(1)=\sqrt{{\cal A}(1)}$, so we can now write $I_\gamma$ as,
\begin{equation}
I_\gamma=
\frac{1}{4}(1- {\cal A}(1)) +\frac{1}{4} G'(0) 
- \frac{1}{4}\left(\sqrt{{\cal A}(1)}-{\cal A}(1)\right)
- \frac{1}{4}\int_0^1 d\lam\,(1-\lam)\,G''(\lam)\,(G(\lam)-1). 
\label{eq:Ifinal}\end{equation}
The term in (\ref{eq:Ifinal}) proportional to $G'(0)$ is nonpositive because
$G'=1/2\,\theta\, G$ is manifestly nonpositive along ingoing null rays.
The next to last term in (\ref{eq:Ifinal}) 
is negative because ${\cal A}(1)<1$. The last term is negative if we assume the
null energy condition,
\begin{equation}
k_a T^{ab}k_b \geq 0, \label{eq:NEC}\end{equation}
because then $G''=-1/2\,(\sigma^2 + 8\pi kTk) G$ is manifestly negative. 
Equation (\ref{eq:fmwproof}) follows.

This demonstrates that the Bousso bound follows from the covariant Bekenstein 
bound with the additional assumption of the null energy condition, 
and also demonstrates under what
conditions the Bousso bound can be saturated.  Namely, to saturate the Bousso bound
in this situaion it is necessary that:
\begin{enumerate} 
\item ${\cal A}(1)=0$.  
\item $\theta|_{\lam=0}=0$.  
\item $\sigma^2 + 8\pi\,kTk$ vanishes if $\lambda\neq 1$ or ${\cal A}(\lam)
\neq 1.$ 
\end{enumerate}
The first requirement implies that the Bousso bound can be saturated only in
its weak form, (\ref{eq:Bousso}), and not its stronger form, 
(\ref{eq:Bousso2}) (except when the cutoff 2-surface $B'$ vanishes so that
(\ref{eq:Bousso2}) and (\ref{eq:Bousso}) are equivalent).
Hence, we consider lightsheets which are
terminated only at caustics.
The second requirement is that the expansion vanish at the lightsheet
boundary $B$, as would be the case for the past directed lightsheet from just inside
the horizon of a black hole.
Notice that the second requirement is a condition on
the choice of the boundary two-surface, and is violated infinitessimally
if a two-surface satisfying the condition is deformed infinitessimally.  
This will be important when
we consider the contribution of shear to the entropy bounds.  
Finally, the third requirement necessitates
that there be no ``source'' of entropy except at points of vanishing
expansion.  This is reminiscent of the membrane paradigm \cite{membrane}, and
also of an operational definition of black hole entropy by Pretorius, Vollick
and Israel (PVI)\cite{pvi}.  PVI define the entropy of a black
hole as the entropy that must be given to a thin shell of matter brought
from infinity 
to its Schwarzschild horizon in order to maintain mechanical and thermal 
equilibrium (with the local acceleration temperature on the shell) during
the process.  It may
be possible to formulate such a definition covariantly making use of these
ideas, although we will not be more precise about such a relation here.
We also point out reference \cite{oppenheim}, where it was also argued that 
a thin spherical shell held in mechanical and thermodynamic equilibrium at its
horizon would have entropy $S_{BH}=A/4$.

\begin{figure}[t]
\epsfxsize=2.5in
\centerline{\epsfbox{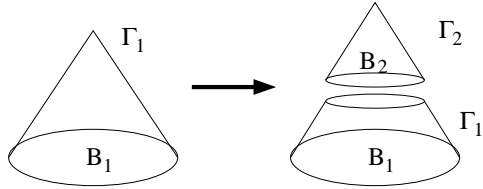}}
\caption{The entropy is additive over the lightsheet, so the lightsheet can be
broken up into sections.}
\label{fig:sections} \end{figure}
\begin{figure}[t]
\epsfxsize=1.5in
\centerline{\epsfbox{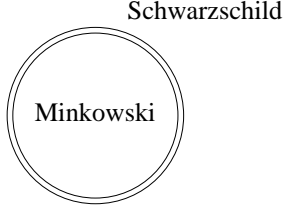}}
\caption{A thin shell of matter separates regions of Schwarzschild and 
flat spacetimes.}
\label{fig:shell} \end{figure}

It is necessary {\em a priori} to distinguish between the Bousso bound 
(\ref{eq:Bousso}) and 
the area law for black holes, \begin{equation}
S_{BH}=A_h/4, \label{eq:BH} \end{equation}
where $A_h$ is the area of the black hole horizon.
In a black hole spacetime we would like it to be the case that an entropy bound
somehow be related to the horizon area $A_h$, 
as opposed to the lightsheet boundary
area $A(B)$.  
Note that the entropy, $S=\int_\Gamma s\cdot k \,d\lam\,d^2x$, can be broken up into
sections as in Figure~\ref{fig:sections}, so that $S=\sum_i S_i$, 
where $S_i$ is the
entropy in the $i$th section, $S_i=A(B_i)\int_{\Gamma_i}s\cdot k\,
{\cal A}(\lam) \,d\lam\,d^2x$.
Intuitively, if matter and black hole horizons are confined to a particular 
$\Gamma_i$ then we would expect the entropy in a lightsheet which contains $\Gamma_i$
to depend only on $\Gamma_i$.  It would be still better if the maximal
entropy satisfying the covariant Bekenstein bound, and 
hence $I_\gamma$ in (\ref{eq:fmwproof}), depended only on $\Gamma_i$.  
If we include the shear in our analysis this is not strictly true, as we will discuss,
but in the absence of shear this result follows immediately from (\ref{eq:fmw1}).
To see how this works explicitly in a specific case, 
consider a static, spherically symmetric, thin shell of matter 
(Figure~\ref{fig:shell}).  
We will assume
the geometry is Minkowski space inside the shell and Schwarzschild outside.
Both geometries have metrics of the form, \begin{equation}
ds^2=f(r)\,dt^2-h(r)\,dr^2-r^2\,d\Omega^2. \end{equation}
We assume the shell of matter is at $r=R$, and we require that the induced
metric be equivalent on both sides of the shell.  For the Schwarzschild region
we have, \begin{equation}
f_{Sch}(r)=1-\frac{2M}{r}, \ \ \ h_{Sch}(r)=1/f_{Sch}(r), \end{equation}
and for the Minkowski region we have, \begin{equation}
f_{Mink}(r)=1-\frac{2M}{R}\equiv f_0,\ \ \ h_{Mink}(r)=1. \end{equation}
The existence of a Killing vector 
$\partial_t$ implies a constant of the motion $e$, such that
\begin{equation}
\frac{dt}{d\lam}=e\,g^{tt}=e/f. \label{eq:kt}\end{equation}
Vanishing of $ds^2$ along the null path then implies that, \begin{equation}
\frac{dr}{d\lam}=\frac{e}{\sqrt{f(r)h(r)}}. \label{eq:kr}\end{equation}
Suppose the spherically symmetric null congruence reaches the shell at $r=R$ when the
affine parameter takes the 
value $\lam=\lam_0$.  Then $r$ goes from $R$ to 0 when $\lam$
goes from $\lam_0$ to 1.  Hence, \begin{equation}
e=\frac{R\sqrt{f_0}}{(1-\lam_0)}. \label{eq:e}\end{equation}
The stress tensor has the form, \begin{equation}
8\pi\,T_a^b =S_a^b\,\frac{\delta(r-R)}{\sqrt{g_{rr}}}. \end{equation}
The Israel junction condition \cite{israel} determines $S_a^b$.  
If $h_{ij}$ is the induced metric on the shell, and $K_{ij}$ is the
extrinsic curvature at the shell, we find that the combination
$K_{ij}-K\,h_{ij}$ (where $K=K_{ij}h^{ij}$) takes the values,
\begin{equation}
K_{ij}-K\,h_{ij}\simeq {\rm diag}\left(\frac{2f}{r\sqrt{h}},\,0,\,\frac{
-r\,(2f+rf')}{2f\sqrt{h}},\,\frac{
-r\,\sin^2(\theta)\,(2f+rf')}{2f\sqrt{h}}\right),\label{eq:K}\end{equation}
in the spherical basis $(t,r,\theta,\phi)$.
The Israel junction condition relates the change in the extrinsic curvature
across the shell to the localized stress tensor on the shell:
\begin{equation}
\Delta(K_{ij}-K\,h_{ij}) = S_{ij}, \end{equation}
where $\Delta(...)$ refers to the difference in the quantity $(...)$ evaluated
just inside and just outside the shell.
Using (\ref{eq:K}) we find,
\begin{equation}
S_t^t=\frac{2}{R}\left(h_{Mink}^{-1/2}(R)-h_{Sch}^{-1/2}(R)\right). 
\label{eq:Stt}\end{equation}
Note also that the shear vanishes on a spherically symmetric lightsheet in a 
spherically symmetric spacetime.
We now have all of the information required to calculate $I_\gamma$ defined in
(\ref{eq:fmwproof}).  The result is,
\begin{eqnarray}
I_\gamma/{\cal A}(\lam_0) &=&\frac{1}{8} \,
\int_0^1d\lam\,8\pi\,k^a T_{ab} k^b =\frac{1}{4}
(1-\sqrt{f_0}) \\
&=&\frac{1}{4}\left(1-\sqrt{1-\frac{2M}{R}}\right). \end{eqnarray}
As expected the Bousso bound is saturated, {\em i.e.} $I_\gamma=1/4$, when the matter shell
approaches its Schwarzschild radius.  Note also that this result is independent
of the size of the boundary of the Bousso lightsheet (as long as it is larger than
the matter shell).  This is consistent with our expectation that only the region 
``containing'' the source of entropy contributes to the entropy in the lightsheet.

In this setting, we can also be more precise as to the 
relation between the original Bekenstein bound and its covariant version.
For the thin static 
shell the entropy is written in terms of the entropy flux vector
as, \begin{eqnarray}
\frac{S}{A}&=&\int d\lam\,s^tk_t \label{eq:S1}\\
&=&\int dr\,\frac{s^tk_t}{k^r}, \label{eq:S2}\end{eqnarray}
where in (\ref{eq:S2}) we used $k^r=dr/d\lam$.  Using (\ref{eq:kt}) and
(\ref{eq:kr}) we can write, \begin{equation}
s^t=\frac{S}{A}\,\frac{1}{\sqrt{f}}\,\frac{\delta(r-R)}{\sqrt{g_{rr}}}.
\end{equation}
In the limiting case that the shell forms a black hole $R=2M$, and 
comparing $S_t^t$ with the black hole energy $E_{BH}=M$, we define
the energy of the shell $E$ via, \begin{equation}
S_t^t=\frac{16\pi E}{A}. \label{eq:Stt2}\end{equation}
With this definition of $E$, and using (\ref{eq:kt}) and (\ref{eq:e}),
the covariant Bekenstein
bound (\ref{eq:fmw1}) can be written exactly as the original Bekenstein
bound (\ref{eq:Bekenstein}).  But notice from (\ref{eq:Stt}) 
that $E\neq M$ except for the case
when the shell is at its Schwarzschild radius.  In fact, from (\ref{eq:Stt2})
we can determine that $E_{fall}\equiv 2E$ 
is the mass as seen by a local freely falling 
observer, so the Bekenstein bound should more precisely be
written as, \begin{equation}
S<\pi E_{fall} R.\end{equation}

Although it is nice to have rederived some familiar results, 
we are immediately led to a puzzle.  What if there was no matter shell?  In 
Schwarzschild spacetime there is no stress tensor, and so no source of entropy if
we assume the covariant Bekenstein bound.  We will see in the next section
that the singularity does not provide an escape to this conclusion, as long as
we assume the covariant Bekenstein bound.  
Then where might the entropy be, and how
is this situation related to the case of a matter shell sitting at its horizon?
We will suggest a solution to this puzzle without denouncing the covariant
Bekenstein bound in the next section.

\section{Shear and gravitational entropy}

At the end of the last section we argued that if we assume a covariant Bekenstein bound
of the form (\ref{eq:fmw1}) then the entropy contained in a spherically symmetric
lightsheet in Schwarzschild spacetime vanishes.  
Thus, the horizon, and all other possible lightsheets
with the same symmetry, will remain shear-free.
One can also imagine
arranging a spherically symmetric impulsive gravitational wave front 
\cite{impulsive}
which passes through the light sheet.  However, this wavefront would also not 
impart any shear to a spherically symmetric null geodesic congruence, and hence could
not contribute any entropy to a spherically symmetric lightsheet by the argument
above.

This would be in contradiction to the intuition that such a
lightsheet which ``contains'' a black hole should measure its entropy.  One might
be concerned that the singularity should somehow 
contribute to the Bousso bound in the
presence of a black hole, but this is not the case.  One can consider a deformation
of the spherically symmetric Bousso
lightsheet which avoids the singularity at all but a finite number of points.

The easiest way to see that the entropy vanishes in the
spherical limit is by using the derivation of the Bousso bound described in the
previous section and in \cite{fmw}.  
For the case of the
spherically symmetric lightsheet we find $G'(0)=-1$, which exactly
cancels the first term in (\ref{eq:Ifinal}). (We assume that ${\cal
A}(1)=0$ on our lightsheet.)  Because $G'(0)$ varies smoothly as the lightsheet
is deformed, it must approach its spherical value of -1 in the spherical limit.
$I_\gamma$ is manifestly positive, so
the last term in (\ref{eq:Ifinal}) must therefore vanish in the spherical limit.
Hence, avoiding the singularity by an
infinitessimal deformation of the lightsheet boundary does not alter
the result that the entropy ``inside'' the black hole vanishes assuming the
covariant Bekenstein bound.
Notice that this analysis relied only
on the behavior of the lightsheet at its boundary, and we were able to eliminate
all reference to the bulk of the lightsheet.

\begin{figure}[t]
\epsfxsize=4.5in
\centerline{\epsfbox{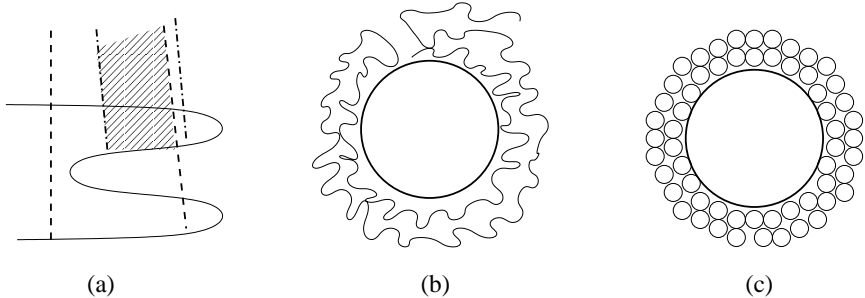}}
\caption{(a) A portion of a folded lightsheet.  The shaded region is overcounted in
the sense described in the text.  (b) A portion of the boundary of a 
``crinkly'' lightsheet surrounding a black hole horizon.  (c) The crinkly lightsheet
is approximated by a ``space-filling'' set of smaller lightsheets.}
\label{fig:fold} \end{figure}

Given this result, 
rather than reject the covariant Bekenstein bound we would like to point out that two
lightsheets which ``contain'' all matter and black holes in a given spacetime need not
observe the same entropy.  In this sense, the entropy of a spacetime is observer
dependent.  Indeed, a generic lightsheet in the Schwarzschild geometry will be sheared,
and may have a nonvanishing entropy according to (\ref{eq:fmw1}).  This line of
thinking begs the question, is there a maximal entropy lightsheet, perhaps with
entropy given by the black hole horizon, in the case of a black hole spacetime? 
In this general sense the answer is clearly no, because even if we only consider
Bousso lightsheets with connected boundaries, 
the lightsheet may fold back on itself (as in 
Figure~\ref{fig:fold}a) and overcount
the entropy in a given spacelike region.  
Hence, a poor choice of Bousso lightsheet
can have as large an entropy as desired, while still satisfying the Bousso bound.
We would like to consider lightsheets which do not overcount
the entropy.  Similar concerns also appear in \cite{low}.
One choice of lightsheet which satisfies this intuitive restriction
is a crinkly lightsheet which wraps back on itself many times, extending out to
scri (Figure~\ref{fig:fold}b).  
The lightsheet can be approximated by a large number of smaller lightsheets,
which are space-filling in the appropriate sense as the lightsheet becomes suitably
crinkled (Figure~\ref{fig:fold}c).  
Alternatively, at least for static spacetimes
we can consider a Bousso lightsheet formed from a 
disconnected set of boundary balls in a space-filling limit.  Perhaps such a 
lightsheet
could have an entropy given by the black hole entropy.  We will argue that,
assuming saturation of the covariant Bekenstein bound,  the entropy
in such a lightsheet indeed scales with the black hole area.  Then we will discuss
the interpretation of this result.

The explicit calculation of $I_\gamma$ for closely packed small lightsheets
depends on two regulators, and potentially the shape of the lightsheet boundaries.
The two regulators are the size of the small lightsheets, 
and how close to the
black hole horizon the lightsheets should be allowed to
probe.  We will choose both regulators
to be Planck size in the static frame.  
For notation, we take the size of the small lightsheets
to be $L$, and we will integrate over such lightsheets to the position
$r=2M+\epsilon$ in Schwarzschild coordinates.  
In that case, as an order of magnitude $k_t\sim L$ because each null ray traverses
a distance on the order of L when the affine parameter goes from 0 to 1. 
In Schwarzschild metric, the tensor 
$B_{ab} \equiv \nabla_a k_b$, which contains the shear, contains the term,
\begin{equation}
B_{tr}\sim \frac{LM}{r^2}\frac{1}{(1-2M/r)}. \end{equation}
The leading term in the shear squared is,
\begin{equation}
(\sigma_{ab})^2\sim B_{tr}B_{tr}g^{tt}g^{rr} = B_{tr}^2. 
\end{equation}
Multiplying this by the number of balls in a shell of radius $r$ gives a factor
of $r^2/L^2$,  and
integrating over the size of a ball gives a factor $L^2$, so  
\begin{equation}
S_{shell} \sim  (\sigma_{ab})^2  \frac{r^2}{L^2} L^2
\sim \frac{L^2M^2}{r^2(1-2M/r)^2}.
\end{equation}
Now we want to integrate $S_{shell}$ out to infinity. If we regulate the lightsheets as
suggested above, then integrating over shells gives , \begin{eqnarray}
S &\sim & \int_{2M+\delta}^\infty \frac{dr}{L}\,S_{shell} \\
  &=& M^2 \,L/\delta. \label{eq:Sest}\end{eqnarray}
Note that we chose to regulate the size of the lightsheets in the static frame,
so that there are no additional metric factors in the integral.
If we choose the two regulators $\delta$ and $L$ to be the Planck length,
then from (\ref{eq:Sest}) it follows that, \begin{equation}
S\sim M^2 \sim A_h. \end{equation}

Notice that the shear (or Weyl) 
contribution to $S$ comes entirely from the region within
a Planck length of the black hole horizon.  We could have chosen the size of the
outermost shell arbitrarily, and as long as it is much larger than the Planck 
scale the resulting entropy in our approximation would be unchanged.  This is
consistent with the intuitive notion that the entropy should be contained within
a thin shell around the horizon (the membrane paradigm).  In this sense, the 
calculation of the entropy for the thin shell of matter in the previous section is
analogous to the calculation in this section of the Weyl entropy.

We have argued that the entropy in a space-filling lightsheet (in the sense
described above) is proportional to the horizon area, but we have not calculated
the coefficient.
It would be nice to understand under which circumstances the black hole entropy
$S_{BH}=A_h/4$ would be obtained.  The dependence of such a result on the shape
of the lightsheets would also be interesting to explore, but is beyond the
scope of this paper.  A numerical exploration of these issues is in progress
\cite{BCE}.

It is worth mentioning that there is another logical possibility concerning the 
entropy contained on this space-filling light sheet:  it could be
the case that this lightsheet measures purely gravitational degrees
of freedom, which sit just outside the horizon, and which have
not been properly included in previous discussions of black hole
entropy.  This interpretation is suggested by the example
of the thin spherically symmetric shell of matter sitting just
at the Schwarzschild horizon (which we considered above).  In addition
to the contribution from the matter shell on a very thin spherically
symmetric Bousso lightsheet, we could include the contributions from
the small, closely packed lightsheets exterior to the shell.  We would
then have two contributions to the entropy, both of which scale precisely
like the area of the horizon.  If this is the correct way to think
about the contribution to the entropy from the small, closely packed
lightsheets, then it suggests a `new' version of the generalized second
law (GSL):  In addition to the usual matter ($S_{matter}$) and horizon
($S_{BH}$) contributions to the total entropy, $S_{Total}$, perhaps we should
also include a purely gravitational term $S_{grav}$:
\begin{equation}
S_{Total} = S_{matter} + S_{BH} + S_{grav}
\end{equation}

The statement of the GSL would then be that $S_{Total}$ can never 
decrease.  Note that this would imply that the process of Hawking 
evaporation does not necessarily generate a huge amount of entropy.
This is because a lot of entropy could already be contained in 
$S_{grav}$, and hence both $S_{grav}$ and $S_{BH}$ could be 
converted to pure $S_{matter}$ (thermal radiation) at the endpoint of
the evaporation process.

\section{Discussion}
We have studied Bousso's covariant entropy bound and its relation to the
covariant Bekenstein bound.  We found that a thin spherically symmetric
shell of matter, which
saturates the covariant Bekenstein bound and sits at its Schwarzschild horizon,
gives rise to the expected black hole entropy on a large spherically symmetric
lightsheet.
We also found that a more fine grained lightsheet which explores the region 
outside
the black hole gives a result proportional to the black hole area, and we conjecture
that an appropriate choice of lightsheet would give the correct
coefficient of 1/4 in the entropy-area relation.  

If this is indeed correct, 
and if we are to interpret this result as due to gravitational entropy
as suggested in the last section, then we are led to a remarkable
conclusion.  In the formation of a black hole by a thin shell of matter, the
question of where the entropy of the black hole is contained is ambiguous.  The
question depends on a choice of lightsheet and is not the same for all 
lightsheets,
even for space-filling lightsheets (when the notion of space-filling is well 
defined).  The black hole entropy can be interpreted either as gravitational
entropy, which is bounded by the shear on fine grained lightsheets; or it
can be interpreted as due to matter entropy in the case of the thin shell 
discussed above.  The latter interpretation is similar to the operational 
definition of entropy given by Pretorius, Vollick and Israel \cite{pvi}, as
discussed in the text.  On the other hand, the gravitational 
interpretation suggests that in order to probe the entropy gravitationally short
distance probes are required, as opposed to the long distance probes which 
measure the matter entropy.  This indicates 
a sort of ultraviolet-infrared duality, although different in nature to the
ultraviolet-infrared duality of \cite{uv-ir}.  

Alternatively, as discussed at the end of the previous section,
it may be the case that the shear entropy through the little lightsheets is to be
interpreted as an addition to the usual black hole entropy.  In this case the 
generalized second law should be further generalized to reflect this gravitational
contribution to the entropy.  It would be interesting to make such a statement more
precise, by finding an appropriate class of lightsheets and studying the time
evolution of the entropy through those lightsheets, assuming the covariant
Bekenstein bound as we did in this paper.  
Even if in a particular time
slicing and a clever choice of lightsheet
a generalized second law could be deduced, the challenge will be making
such a statement generally covariant.

To be fair, we have not precisely calculated the contribution of the shear
to the Bousso bound, but only argued that this contribution is proportional to
the area of the black hole horizon.  It would be interesting to do a more
explicit calculation, and also to study the effect of modifying the shape and
size of the fine-grained lightsheets.  There are many ``derivations'' of the
black hole entropy law and various formulations of entropy bounds in the
literature.  Most of them are not covariant.  It is necessary to compare
older approaches to black hole entropy to modern covariant approaches in
the hope of better understanding gravity.  Much remains to be done in this
regard.

In addition, it is worth commenting that there is a nice separation of ``quantum''
and ``gravitational'' effects in the covariant Bekenstein bound.  Putting
$\hbar$ and $G$ back into (\ref{eq:fmw1}) gives, \begin{equation}
\hbar\,
s_\Gamma\cdot k\leq (1-\lambda) (\pi\, k_a T^{ab}k_b + G\,\sigma_{ab}\sigma^{ab}/8).
\end{equation}
Although the parameters $\hbar$ and $G$ are dimensionful and can be rescaled to
one,
it is tempting to interpret the covariant entropy bound
as due to a purely quantum mechanical constraint on the entropy
of matter and a quantum gravitational constraint with regards to the gravitational
Weyl entropy.  As there is no $G$ appearing in the part of the covariant
Bekenstein bound related to the stress tensor, the entropy-area relation we found for
the spherical matter shell relies on gravity only clasically.  This is similar
in spirit to previous studies of spherical shells \cite{pvi,oppenheim}.
We also note that such a separation of 
$\hbar$ and $G$ does not follow from the second set of constraints
under which the Bousso bound has been proven, (\ref{eq:fmw2}).

\section*{Acknowledgments}
We thank Raphael Bousso, Gary Gibbons, 
Ted Jacobson, Andreas Karch, Ami Katz, Matt Strassler, Robert Wald, and Neal Weiner 
for useful discussions.  We are especially grateful to Tanmoy Bhattacharya for
collaboration on much of this project.  AC is supported by a PPARC Advanced Research 
Fellowship at Queen Mary, University of London. JE is supported by the DOE under
contract DE-FGO3-96-ER40956.

\vspace{\baselineskip}

\end{document}